\documentclass[12pt]{article}

\usepackage{microtype}
\usepackage{graphicx}
\usepackage{subfigure}
\usepackage{booktabs} % for professional tables
\usepackage{url}

\usepackage{hyperref}
\usepackage{breakurl}

\usepackage[accepted]{icml2021}

\icmltitlerunning{Insuring Uninsurable Risks from AI: The State as Insurer of Last Resort}

\begin{document}
\thispagestyle{plain}
\pagestyle{plain}

% \twocolumn[
\icmltitle{Insuring Uninsurable Risks from AI: Government as Insurer of Last Resort}



\icmlsetsymbol{equal}{*}

\begin{icmlauthorlist}
\icmlauthor{Cristian Trout}{ind}
\end{icmlauthorlist}

\icmlaffiliation{ind}{Independent Researcher, Cambridge Boston Alignment Initiative}

\icmlcorrespondingauthor{Cristian Trout}{ctroutcsi@gmail.com}

\icmlkeywords{AI Policy, Liability, Insurance, Mechanism Design, ICML}

% ]



\printAffiliationsAndNotice{} % otherwise use the standard text.

\begin{abstract}
\textit{Many experts believe that AI systems will sooner or later pose uninsurable risks, including existential risks. This creates an extreme judgment-proof problem: few if any parties can be held accountable \textit{ex post} in the event of such a catastrophe. This paper proposes a novel solution: a government-provided, mandatory indemnification program for AI developers. The program uses risk-priced indemnity fees to induce socially optimal levels of care. Risk-estimates are determined by surveying experts, including indemnified developers. The Bayesian Truth Serum mechanism is employed to incent honest and effortful responses. Compared to alternatives, this approach arguably better leverages all private information, and provides a clearer signal to indemnified developers regarding what risks they must mitigate to lower their fees. It's recommended that collected fees be used to help fund the safety research developers need, employing a fund matching mechanism (Quadratic Financing) to induce an optimal supply of this public good. Under Quadratic Financing, safety research projects would compete for private contributions from developers, signaling how much each is to be supplemented with public funds.}
\end{abstract}

\section{Background}

Many experts believe AI systems will, sooner or later, pose uninsurable risks, including existential risks \citep{grace_thousands_2024,bengio_managing_2024}. If so, it will be impossible to hold accountable the parties liable for such harms (or their insurers).

Weil \citeyearpar{weil_tort_2024} proposes to solve this extreme judgment proof-problem by assigning punitive damages to harms that are \textit{correlated} with uninsurable risks (where the correlation would be estimated by courts and juries). While of interest, this solution has several problems. First, is it's novelty: this would be an unprecedented application of punitive damages that may violate the Due Process Clause \citeyearpar[40-44, 50-53]{weil_tort_2024}, requiring a major doctrinal shift that would cut across all of tort law. Second, correlates of uninsurable risks might be difficult to find. Third, given the high uncertainty involved, correlation estimations by courts will likely be \textit{ad hoc}, high variance, and fail to leverage all available information. Fourth and finally, punitive damages for correlated risks will send a very oblique and noisy signal to liable parties: its effectiveness at actually inducing greater care taken is doubtful. Liable parties might find powerful legal teams to be a safer investment than investments in safety.

Historically, the solution to uninsurable (albeit, non-existential risks) has been for government to step into its role as insurer of last resort \citep{moss_when_2004}, as seen in government provided reinsurance for terrorism risk insurance \citep{federal_insurance_office_report_2022} or indemnification schemes for nuclear power operators \citep[sec. 3.2]{united_states_nuclear_regulatory_commission_price-anderson_2021}. Such programs can be in the government’s interest for several reasons. First, by creating a more predictable legal environment and making insurance more affordable, they spur the economy (or particular industry) in the short-term while protecting it against future shocks by increasing insurance uptake \citeyearpar[sec. 1.1]{united_states_nuclear_regulatory_commission_price-anderson_2021}\citep[6, 7]{michel-kerjan_how_2006}\citep[178]{hubbard_economic_2005}. Second, by encouraging or mandating \textit{ex ante} contributions from the private sector, governments can lower their financial exposure to risk \citep{carroll_assessing_2004}. Governments cannot credibly commit to \textit{not} bail out a critical economic sector or \textit{not} provide relief to victims in the event of a major disaster: the government is always implicitly exposed to such risk.\footnote{This financial reason, based on \textit{ex post} costs to governments, will not apply to existential risks: governments are also judgment-proof in the face of such risks.} Third, governments might mitigate the moral hazard it generates as implicit insurer (or lender) of last resort (see e.g. the ``Too Big To Fail" effect \citep{strahan_too_2013}).

In the context of solving the judgment-proof problem, this last reason is the most interesting. While such programs might reduce moral hazard over the baseline (no program), current programs can only have done so in a crude manner due to their crude pricing. For example, in the commercial nuclear power case, indemnity fees were charged per plant and set by simply multiplying the maximum power output of said plant by fixed multipliers \citep[sec. 3.2]{united_states_nuclear_regulatory_commission_price-anderson_2021}. This cannot have encouraged operators to take greater care along any dimension other than their choice of maximum power output at the initial design stage.

Risk-based pricing is wanted. A government agency could make risk estimates, but this would be costly and the agency would struggle to collapse the information asymmetry between itself and the well-resourced private actors it insures. This paper proposes another solution, leveraging advances in mechanism design: a survey mechanism, the Bayesian Truth Serum \citep{prelec_bayesian_2004}, could be used to reliably extract and aggregate honest risk estimates from experts, \textit{including the parties insured}. This, it’s argued, better leverages all available information than other solutions.

It’s further argued that if accurate risk-based pricing can be had, a government indemnification program is preferable over Weil's punitive damage regime for producing less litigation and more robustly signaling to insureds of what risks they must mitigate. From the insured’s perspective, it would also more consistently transform large and uncertain \textit{ex post} costs into manageable and certain \textit{ex ante} costs.

Finally, this paper proposes using a contribution matching mechanism, Quadratic Financing \citep{buterin_flexible_2019}, to redistribute collected fees \textit{back} to industry, funding the research required to reduce its uninsurable risks.

Mechanism design has long been recognized as a tool for governance (see e.g. the FCC's auctioning of the electromagnetic spectrum \citep{zaretsky_auctions_1998}), but has seen few sophisticated applications (to this author's knowledge). As a contribution to the literature on regulatory design, this paper hopes to mark out fertile ground for regulatory innovation at the intersection of tort law and mechanism design, resulting in a governance regime distinct from traditional command and control regulation, performance-based regulation, or a regime reliant solely on tort law.

\section{A Risk-Priced Indemnification Program}

An indemnification program is preferred over reinsurance as this removes the intermediary of insurers, allowing the government to directly manipulate incentives of risk-generating parties. Elsewhere I’ve argued that it’s developers (e.g. OpenAI) who should be strictly and exclusively liable for said risks, largely based on their being least-cost avoiders \citep{trout_liability_2024}.

I'll call this proposed scheme the AI Disaster Insurance Program (AIDIP). Participation would be mandatory for developers of AI models trained over a certain effective compute\footnote{Effective compute = FLOPs * an algorithmic efficiency factor, as estimated by a government agency such as the U.S. Artificial Intelligence Safety Institute.} threshold. The core of the program is a risk-priced indemnity fee that developers must pay per training run. It’s recommended the fee be a function of effective compute.\footnote{Cf. the industry’s currently voluntary ``responsible scaling policies" e.g. \citep{anthropic_anthropics_2023}.}

\begin{figure}[h]
\centerline{\includegraphics[width=1\columnwidth]{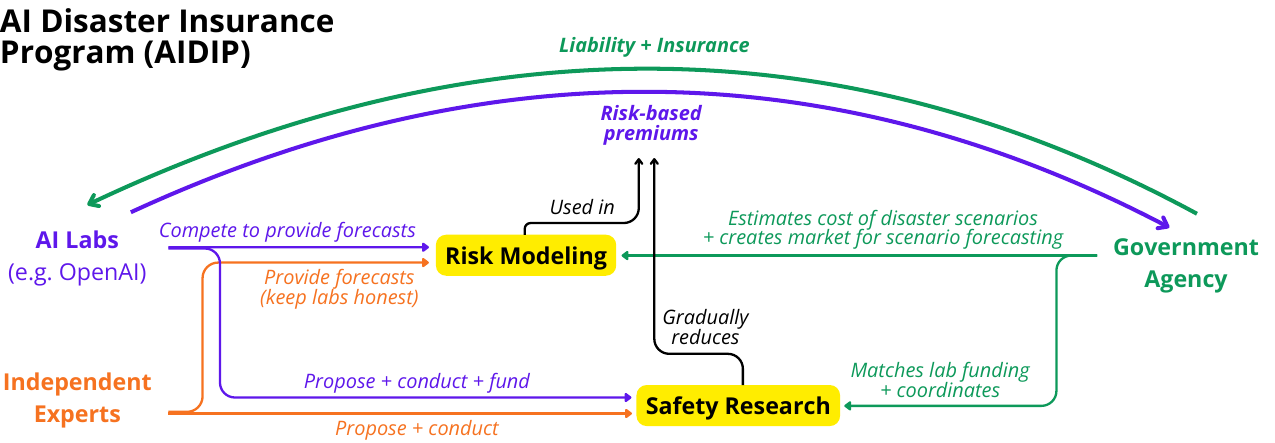}}
\caption{The AI Disaster Insurance Program in schematic form.}
\label{fig: AIDIP}
\end{figure}

A government agency (such as the Department of Homeland Security) would estimate the disutility of various disaster scenarios, but risk-estimation would rely on a survey of public and private experts, \textit{including indemnified developers}. The Bayesian Truth Serum (BTS) is employed to incentivize \textit{effortful} and \textit{honest} risk-estimations from respondents \citep{prelec_bayesian_2004}. BTS rewards responses with high information scores – i.e. responses that are surprisingly common relative to respondents’ predictions of how other respondents will respond. Scaling the BTS payout incents greater effort in information gathering. Honest reporting is a Bayes-Nash equilibrium under BTS – i.e. absent other incentives, a respondent will maximize their expected payout by reporting honestly \textit{if} they believe a large enough majority of other respondents will also report honestly.

A developer (who must pay the fee) obviously has a conflicting incentive to lie (underreport the risks), and can expect other developers to lie. This conflict of interest can be overcome by dramatically scaling BTS’ payout or potentially removed entirely by silencing the developer’s risk-estimation when their \textit{individualized} fee is calculated. (This second option puts developers in a prisoner’s dilemma: they could lower their fees by coordinating, but it’s individually rational to defect, increasing fees for one’s competitors.) An expectation of overwhelming honesty can be created by ensuring the vast majority of respondents are \textit{not} developers but instead independent experts with no conflicts of interest. Where there are no conflicts of interest, relevant insurers and government agencies (such as the newly formed U.S. Artificial Intelligence Safety Institute) could also be respondents.

The survey should be run at regular intervals (e.g. yearly), with the fee scale fixed for that interval. Any developer who wants to train a new frontier AI model during the current window must have participated in the last survey. A small discount on the fee could be offered for having participated in the last several surveys. The government sets the industry's agenda in its choice of survey questions (e.g. “For an AI trained on compute \textit{x}, what’s the likelihood of disaster \textit{D} within time frame \textit{t}?”), clearly signaling to indemnifieds what risks they must mitigate to lower their fees.

To defend against BTS’ (not unusual) vulnerability to collusion, collusion should be heavily fined; whistleblowers, modestly rewarded. Participants would also have to be barred from conducting their own surveys of experts just before the government survey, lest they short-circuit BTS.\footnote{For BTS to reliably induce honest reporting, it's critical that participants' estimates of how other participants will respond be based on a participant's private information regarding the topic in question, not a recent survey of the other participants.}

Using BTS to solve the information asymmetry between the government and indemnifieds has several advantages over relying heavily on government risk estimates and inspecting indemnifieds. It should be cheaper, incenting parties to compete to provide the most informative risk-estimates, all while more reliably aggregating a wider range of private information. It should also be more secure – developers can divulge the \textit{risk implications} of their private information \textit{without} exposing security-sensitive information. Finally, it creates a more cooperative relationship between developers and the government, lending the regime greater legitimacy.

\section{Quadratic Financing for Safety Research}

It's recommended that revenue from the insurance program be used to fund AI related programs in the public's interest. One such program should aim to directly help developers shoulder the cost of the Safety Research (SR) they need to reduce their fees.

Because SR is a public good, developers face a coordination problem. Note that the coordination problem for supplying SR is greatly simplified by the liability and indemnification regime: instead of countless potential victims needing to coordinate, only \textit{developers} need coordinate. To help them solve their coordination problem, ensure an optimal supply of SR, and contribute its own fair share of funds, it's recommended the government employ a fund-matching mechanism, Quadratic Finance (QF), designed to achieve all the above \citep{buterin_flexible_2019}.

Under QF, developers and partnered research institutions would propose various SR projects. Developers then choose to fund to whatever extent whichever projects it likes, knowing the government will top-up a project's total funds according to the QF formula. This top-up scales quadratically in the number of contributors to a project. As with BTS, QF will require basic defenses against collusion (multiple private contributors funding each other in order to receive a higher top-up) and fraud (one private contributor pretending to be multiple).

Projects would essentially be competing for private contributions, signaling where to send public funds. Because of the agenda setting achieved by the liability and insurance program, projects would require minimal vetting. The market then determines which projects achieve that agenda most effectively.

\section{Closing Remarks and Further Research}

This paper proposes internalizing the negative externalities of uninsurable risks from AI through a risk-priced government insurance program (or from another angle, a Pigouvian tax). Revenue from this program then flows back to industry through an SR funding program. 

The overall regime is market-based in that it has private actors compete to provide the most well-informed risk-estimations, and the most effective research projects to reduce said risks. This approach, it’s argued, is cheaper, more responsive to new information, and more effective at protecting the public than alternative solutions to this extreme judgment-proof problem. 

While this paper focuses on elaborating the details of the market mechanisms core to this novel governance regime, it should be emphasized that the robust liability channeling, as discussed further in Trout \citeyearpar{trout_liability_2024}, is no less critical. Without channeling responsibility onto a few, well-resourced, and well-informed private actors these market mechanisms would likely be much less reliable (for appearing less legitimate, being costlier to administer and police, and returning a noisier signal).

As with quasi-regulation via insurance \citeyearpar{trout_liability_2024}, the goal here is not to fully substitute for regulation, but rather to produce effective risk-modeling and safety design for this emerging technology. Once available, well-calibrated regulation is much easier to develop. 

As was done with the Terrorism Risk Insurance Program and Price-Anderson Act, this indemnification and funding program should include sunset mechanisms (an expiration date and or mechanisms for making private actors take ever greater responsibility for managing the risks of their activities, as this becomes possible). This would help ensure the program doesn't outlive its utility and is iterated on to meet changing needs.

While confident in the theoretical soundness of its claims, the paper acknowledges the need for further empirical research into the effectiveness of BTS and QF. Available studies align with theoretical expectations \citep{weaver_creating_2013,pasquini_quadratic_2022}, but more tests are needed.

\section*{Acknowledgements}
This work was supported by the Cambridge Boston Alignment Initiative. Thanks to Andy Haupt, Thomas Larsen, and Mackenzie Arnold for helpful guidance early on. Thanks to Mackenzie Arnold, Gabriel Weil, Trevor Levin, Janet Egan, and Ketan Ramakrishnan for helpful feedback. Special thanks to Warren Zhu for extensive help working through various mechanism design questions.


\bibliography{references}
\bibliographystyle{icml2021}

\end{document}